\documentclass[conference, 10pt]{IEEEtran}
\usepackage[all=tight, sections=normal, tracking=normal, title=normal, mathdisplays=normal, lists = normal, margins = normal]{savetrees}

\ifCLASSINFOpdf
   \usepackage[pdftex]{graphicx}
\else
\fi
\usepackage[cmex10]{amsmath}
\usepackage{amsfonts}
\usepackage{cases}
\newcommand{\inner}[2]{\langle #1 , #2 \rangle}
\newcommand{\comment}[1]{}
\usepackage{xcolor}
\usepackage{comment}
\newcommand{\mc}[1]{{#1}} %relecture par mc (matthieu crussi\E8re)
\newcommand{\mroy}[1]{{#1}} %relecture par mroy (Matthieu ROY)
\newcommand{\commentaire}[1]{{}} 
\makeatletter

\let\origsection\section

\renewcommand\section{\@ifstar{\starsection}{\nostarsection}}

\newcommand\nostarsection[1]{\sectionprelude\origsection{#1}\sectionpostlude}

\newcommand\starsection[1]{\sectionprelude\origsection*{#1}\sectionpostlude}

\newcommand\sectionprelude{
	\vspace{-0.5em}	
}
\newcommand\sectionpostlude{
	\vspace{-0.5em}	
}

\let\origsubsection\subsection
\renewcommand\subsection{\@ifstar{\starsubsection}{\nostarsubsection}}

\newcommand\nostarsubsection[1]{\subsectionprelude\origsubsection{#1}\subsectionpostlude}

\newcommand\starsubsection[1]{\subsectionprelude\origsubsection*{#1}\subsectionpostlude}

\newcommand\subsectionprelude{
	\vspace{-0.4em}	
}
\newcommand\subsectionpostlude{
	\vspace{-0.4em}	
}
\makeatother

\begin{document}
\title{MIMO Channel Hardening:\\ A Physical Model based Analysis}

\author{\IEEEauthorblockN{Matthieu Roy$^\dagger$, St\'ephane Paquelet$^\dagger$, Luc Le Magoarou$^\dagger$, Matthieu Crussi\`ere$^\ddagger$}
\IEEEauthorblockA{$^\dagger$b\raisebox{0.2mm}{\scalebox{0.7}{\textbf{$<>$}}}com Rennes, France\\
$^\ddagger$Univ Rennes, INSA Rennes, IETR - UMR 6164 F-35000 Rennes, France}}
%\and
%\IEEEauthorblockN{St\'ephane Paquelet}
%\IEEEauthorblockA{b\raisebox{0.2mm}{\scalebox{0.7}{\textbf{$<>$}}}com\\ Rennes, France}
%\and
%\IEEEauthorblockN{Luc Le Magoarou}
%\IEEEauthorblockA{b\raisebox{0.2mm}{\scalebox{0.7}{\textbf{$<>$}}}com\\ Rennes, France}
%\and
%\IEEEauthorblockN{Matthieu Crussi\`ere}
%\IEEEauthorblockA{Univ Rennes, INSA Rennes IETR - UMR 6164\\ F-35000 Rennes, France}}


%Telephone: (800) 555--1212, Fax: (888) 555--1212}


\newcommand{\CN}{(i)}

\maketitle

\begin{abstract}
%\boldmath
  In a multiple-input-multiple-output (MIMO) communication system, the multipath fading is averaged
  over {radio} links. {This well-known \emph{channel hardening} phenomenon plays a central role in
    the design of massive MIMO systems. The aim of this paper is to study channel hardening using a
    physical channel model in which the influences of propagation rays and antenna array topologies
    are highlighted. A measure of channel hardening is derived through the coefficient of variation
    of the channel gain.  Our analyses and } {closed form} {results based} on the \mc{used} physical
  model are consistent with those of the literature relying on more abstract \mroy{Rayleigh fading}
  models, but offer further insights on the \mc{relationship with} channel characteristics.
\end{abstract}



% \fi
\IEEEpeerreviewmaketitle

\section{Introduction}
%Multi-antenna techniques have been studied extensively over the last decades. They have been
\mc{Over the last decades, multi-antenna techniques} have been identified as key technologies to improve
the throughput and reliability of future communication systems. They offer a potential massive
improvement of spectral efficiency over classical SISO (single-input-single-output) \commentaire{communication
devices}\mc{systems} {proportionally to the number of involved antennas}. This promising gain has been quantified
in terms of capacity in the seminal work of Telatar \cite{telatar_capacity_1999} and {has recently
  been even more emphasized with the \mc{newly introduced} massive MIMO paradigm \commentaire{systems
  }\cite{marzetta_fundamentals_2016}.}

Moving from SISO to MIMO, the reliability of communication systems improves
tremendously. %The reliability improvement of MIMO systems compared to classical systems is tremendous.
On the one hand in SISO, the signal is emitted from one single antenna and captured at the receive
antenna as a sum of constructive or destructive echoes. \mc{This results in fading effects} leading
to a \mc{potentially} very unstable signal to noise ratio (SNR) depending on the richness of the
scattering environment. On the other hand in a MIMO system, with appropriate precoding, small-scale
multipath fading is averaged over the multiple {transmit and receive} antennas. This
\commentaire{leads to a potentially}\mc{yields} a strong reduction of the received power
fluctuations, hence the channel gain becomes \mroy{locally deterministic} essentially driven by its
large-scale properties.  This effect, sometimes referred to as \emph{channel hardening}
\cite{hochwald_multiple-antenna_2004} has recently been given a {formal} definition based on the
channel power {fluctuations} \cite{ngo_no_2017}.  Indeed, studies on the stability of the SNR are
essential to the practical design of MIMO systems, in particular on scheduling, rate feedback, channel
coding and modulation dimensioning \cite{marzetta_fundamentals_2016, hochwald_multiple-antenna_2004,
  bjornson_random_2016}.  From the definition in \cite{ngo_no_2017}, we propose in this paper a
comprehensive study on channel hardening through a statistical analysis {of received} power
variations {derived} from the propagation characteristics {of a generic ray-based spatial channel
  model}.

\noindent{\bf Related work.} {Channel hardening, measured as the channel gain variance, has recently been studied from several points of view.} 
%Recently some papers studied channel hardening by various means. Those studies used the variance of the channel gain as a measure of channel hardening.
The authors in \cite{martinez_massive_2016} used data from measurement campaigns and extracted the
variance of the received power. A rigorous definition of channel hardening was then given %defined
in \mroy{the seminal work} \cite{ngo_no_2017} based on the asymptotic behavior of the channel
gain %normalized variance
for large antenna arrays. This definition was applied to pinhole channels, i.i.d. correlated and
uncorrelated Rayleigh fading models \cite{bjornson_massive_2017}.  \commentaire{Such models provide
  abstract visions of the physical channel use limiting assumptions
  \cite[p. 225]{bjornson_massive_2017}. %harder.
  Moreover such models cannot directly take into account the shapes of the antenna arrays %directly
  and have to relying on correlation matrices.}
{%{Last but not least,} most of those studies focus on asymptotic cases with infinite antenna arrays.
}

%The capacity of a MIMO system \cite{telatar_capacity_1999} is given by the formula (\ref{form_capacity}), where $P_t$ is the total emitted power, $N_0$ is the noise power, $\mathbf{H}$ is the channel matrix and $\mathbf{\bar{Q}}$ is the input correlation matrix (precoding). The capacity is the maximum achievable rate of communication between the emitter and the receiver. It is the main measure of performance of a communication system.
%
%\begin{equation}
% C = \operatorname{log}_2 ( det (\mathbf{I}_{N_t} + \frac{P_t}{N_0} \cdot \Vert \mathbf{H} \Vert_F^2  \mathbf{\bar{Q}} \mathbf{\bar{H}}^H \mathbf{\bar{H}} )) \quad bps / Hz 
% \label{form_capacity}
%\end{equation}
% 
%In the formula (\ref{form_capacity}) the channel matrix and the precoding matrix are normalized hence the $\Vert \mathbf{H} \Vert_F^2$ normalization term. The term $\rho = \frac{P_t}{N_0} \cdot \Vert \mathbf{H} \Vert_F^2$ is the SNR power at the receiver \cite{loyka_physically-based_2009}. As a conclusion the received power also depends on the channel characteristics. 

%The concept of channel hardening is simple to understand. However, it is hard to find a quantitative measure of it in the literature. It was first described by the variance of the mutual information \cite{hochwald_multiple-antenna_2004}, Then by the asymptotic behavior of the singular values of the channel matrix \cite{narasimhan_channel_2014}. Recently a precise definition of channel hardening based on the variations of the channel power was given in \cite{ngo_no_2017}. In this paper we follow the philosophy of this definition.

\noindent{\bf Contributions.} \mroy{Complementary to this pioneer work, }\commentaire{In this paper
}we propose a non-asymptotic analysis of channel hardening, {{as well as }%and propose
  new derivations of the coefficient of variation of the channel not limited to {classically
    assumed} Rayleigh fading models. % as assumed in many works.
  Indeed, channel hardening is analyzed herein using a physically motivated ray-based} channel model
widely used in wave propagation. Our approach is consistent with previous {studies % on the subject
}\cite{ngo_no_2017, martinez_massive_2016}, but gives deeper insights on channel hardening. In
particular, we managed to provide an expression of the channel hardening measure in which the
contributions of the transmit and receive antenna arrays, and the propagation conditions can easily
be identified, and thus interpreted.

\noindent\textbf{Notations.} Upper case and lower case bold symbols are used for matrices and
vectors. {$z^*$ denotes the conjugate of $z$}. $\vec{u}$ stands for a three-dimensional (3D) vector. $\inner{.}{.}$ and $\vec{a}
\cdot \vec{u}$ denote the inner
product between two vectors of $\mathbb{C}^N$ and 3D vectors, respectively.
								 %, $\vec{a}
                                %\cdot \vec{u}$ designates the inner product between two 3D
                                %vectors. 
$[\mathbf{H}]_{p, q}$ is the element of matrix $\mathbf{H}$ at row $p$ and column $q$.
$\Vert \mathbf{H} \Vert_F$, $\Vert \mathbf{h} \Vert$ and
$\Vert \mathbf{h} \Vert_p$ stand for the Frobenius norm, the  euclidean norm and the p-norm,
respectively.  $\mathbf{H}^H$ and $\mathbf{H}^T$ denotes the conjugate transpose and the transpose
matrices.  {$\mathbf{\bar{H}}$ denote the normalized matrix $\mathbf{H}/\vert \vert \mathbf{H} \vert \vert_F$.} $\mathbb{E}\left\{ . \right\}$ and $\operatorname{\mathbb{V}\text{ar}}\left\{ .  \right\}$
denote the expectation and variance.

\section{Channel Model}
We consider a narrowband MIMO system (interpretable as an OFDM subcarrier) with $N_t$ antennas at the transmitter and $N_r$ antennas at
the receiver, such that
\begin{equation*}
\mathbf{y} = \mathbf{H}\mathbf{x} + \mathbf{n},
\label{narrow_band_cm}
\end{equation*}
{with $\mathbf{x} \in \mathbb{C}^{N_t \times 1}$, $\mathbf{y} \in \mathbb{C}^{N_r\times 1}$ and
  $\mathbf{n} \in \mathbb{C}^{N_r\times 1}$ the vectors of transmit, receive and noise samples,
  respectively.} $\mathbf{H}\in \mathbb{C}^{N_r \times N_t}$ is the MIMO channel matrix, {whose
entries $[\mathbf{H}]_{i, j}$ are the complex gains} of the SISO links between transmit antenna
$j$ and receive antenna $i$. %The overall gain of the MIMO channel is the sum of all the SISO channel gains $\Vert [\mathbf{H}]_{i, j} \Vert^2$, and can be expressed as
{The capacity of the MIMO channel can be expressed as \cite{telatar_capacity_1999}
%\begin{equation*}
%\Vert \mathbf{H} \Vert^2_F = \sum_{i, j} \Vert [\mathbf{H}]_{i, j} \Vert^2.
%\label{hnorm}
%\end{equation*}

\begin{equation}
C = \operatorname{log}_2 ( \operatorname{det} (\mathbf{I}_{N_t} + \rho \mathbf{\bar{Q}} \mathbf{\bar{H}}^H \mathbf{\bar{H}} )) \quad bps / Hz, 
\label{capa}
\end{equation}
where $\rho = \frac{P_t}{N_0} \vert \vert \mathbf{H} \vert \vert_F^2$  
with $\mathbf{\bar{Q}}\in \mathbb{C}^{N_t \times N_t}$, $P_t$ and $N_0$ the input correlation matrix (precoding), emitted power and noise power.}  
$C$ is a monotonic function of the optimal received SNR $\rho$ \cite{loyka_physically-based_2009}, hence $\Vert \mathbf{H} \Vert^2_F$ directly influences the capacity of
the MIMO channel. It is then of high
interest studying the spatial channel gain variations to predict the stability of the capacity.

{In the sequel, we will consider that the channel matrix $\mathbf{H}$ is obtained from the following}
generic multi-path 3D ray-based model considering planar wavefronts \cite{raghavan_sublinear_2011, zwick_stochastic_2002}, \cite[p. 485]{bjornson_massive_2017}
\begin{equation}
\mathbf{H}(f) =  \sqrt{N_tN_r}\sum_{p = 1}^{P} c_{p}  \mathbf{e_r}(\vec{u}_{rx, p})\mathbf{e_t}(\vec{u}_{tx, p})^H.
\label{generic_cm}
\end{equation} 
Such channel consists of a sum of $P$ physical paths where $c_{p}$ is the complex gain of path $p$
\mc{and} $\vec{u}_{tx, p}$ (resp. $\vec{u}_{rx, p}$) \mc{its} {direction of departure - DoD - (resp. of arrival
- DoA -). In  \eqref{generic_cm} }$\mathbf{e_t}$ and $\mathbf{e_r}$ are the so-called
{\textit{steering vectors}} {associated} to the transmit and receive arrays. They contain the path
differences of the plane wave from one antenna to another and are defined as \cite{zwick_stochastic_2002}
\begin{equation}
\mathbf{e_t}(\vec{u}_{tx, p}) = \frac{1}{\sqrt{N_t}}\left[\operatorname{e}^{2 j \pi \frac{\vec{a}_{tx, 1} \cdotp \vec{u}_{tx, p} }{\lambda}}, \cdots, \operatorname{e}^{2 j \pi \frac{\vec{a}_{tx, N_t} \cdotp \vec{u}_{tx, p} }{\lambda}}   \right]^T,
\label{steering_vect}
\end{equation}
{and similarly for} $\mathbf{e_r}(\vec{u}_{rx, p})$. The steering vectors depend not only on the
DoD/DoA of the impinging rays, but also on the topology of the antenna arrays. The latter are
defined by the sets of vectors $\mathcal{A}_{tx} = \left\{\vec{a}_{tx, j} \right\}$ and
$\mathcal{A}_{rx}= \left\{\vec{a}_{rx, j} \right\}$ {representing the positions of the antenna
elements in each array given an arbitrary reference.}

Such channel model has already been widely used (especially in its 2D version) {%and can also be found in
}\cite{raghavan_sublinear_2011, zwick_stochastic_2002}, {%It has been
}verified through measurements \cite{samimi_28_2016} for millimeter waves {and studied in the context of channel estimation
\cite{magoarou_parametric_2017}.} {In contrast to Rayleigh channels, it explicitly takes into account the propagation conditions and the topology of the antenna arrays.% Contrary to the common i.i.d. Gaussian model discussed above, the propagation conditions and the topology of the antenna arrays are included in this model.
}
%Such a general channel model has also been studied in the context of channel estimation\cite{magoarou_parametric_2017}.

%$\mathbf{n} \sim \mathcal{C}\mathcal{N}(0, \mathbf{I})$

{In the perspective of the following sections, }let
$\mathbf{c} = \left[ \vert c_1 \vert, \cdots ,\vert c_P \vert \right]^T$ denote the vector
consisting of the amplitudes of the rays. \commentaire{Then, }$\Vert \mathbf{c} \Vert^2$ is the \commentaire{total} aggregated
power from all rays, %, that is the overall large-scale power. Note that such term 
 corresponding to
large-scale fading due to path-loss and shadowing.
\label{channel_model}

 %$\vec{u}_{rx, p}$ and $\vec{u}_{tx, p}$ direction of arrival and departure of the 3D rays.
 %General channel model that can be used for 3D antenna modelisation. Sparce MIMO channel.
 %Steering vectors defined by : $\mathbf{e_t}(\vec{u}_{tx, p}) = \frac{1}{\sqrt{N_t}}\left[\operatorname{e}^{2 j \pi \frac{\vec{a}_{tx, 1} \cdotp \vec{u}_{tx, p} }{\lambda}}, \cdots, \operatorname{e}^{2 j \pi \frac{\vec{a}_{tx, N_t} \cdotp \vec{u}_{tx, p} }{\lambda}}   \right]$, same for $ \mathbf{e_t}(\vec{u}_{tx, p})$

{\section{Channel Hardening}}

%\subsection{Definition}
{\noindent {\bf Definition.}} Due to the multipath behavior of  propagation channels, classical SISO systems suffer from a strong
fast fading phenomenon at the scale of the wavelength resulting in strong capacity fluctuations (\ref{capa}). MIMO systems average the fading phenomenon over the antennas so that the
channel gain varies much more slowly. This effect is called \textit{channel hardening}. In this paper, the relative variation of the channel gain $\Vert \mathbf{H} \Vert^2_F$, {called} coefficient of variation ($CV$) is evaluated to quantify the channel
hardening effect as previously introduced in \cite{ngo_no_2017, bjornson_massive_2017}:
\begin{equation}
CV^2 = \frac{\mathbb{V}\text{ar} \left\{ \Vert \mathbf{H} \Vert^2_F \right\} }{\mathbb{E} \left\{ \Vert \mathbf{H} \Vert^2_F \right\}^2} = \frac{\mathbb{E} \left\{ \Vert \mathbf{H} \Vert^4_F \right\}-\mathbb{E} \left\{ \Vert \mathbf{H} \Vert^2_F \right\}^2}{\mathbb{E} \left\{ \Vert \mathbf{H} \Vert^2_F \right\}^2}
\label{non_asymt_ch}
\end{equation}
\mc{In \eqref{non_asymt_ch}} the statistical means are obtained upon the model which govern the entries of $\Vert \mathbf{H} \Vert^2$ given random positions of the transmitter and the receiver. % is a function of multiple random variables specified by a statistical model. 
This measure was previously  applied to a $N_t \times 1$ correlated Rayleigh channel model
$\mathbf{h}\sim \mathcal{CN}(\mathbf{0},\mathbf{R})$ \cite{ngo_no_2017}, \cite[p. 231]{bjornson_massive_2017}. %with {covariance matrix $\mathbf{R}= \mathbb{E}\left[{\mathbf{h}\mathbf{h}^H}\right]$}. 
{In that particular case, \eqref{non_asymt_ch} becomes%The result with this particular model is given by
\begin{equation}
	CV^2 = \frac{\mathbb{E}\left\{ \vert \mathbf{h}^H\mathbf{h} \vert^2 \right\} - \operatorname{Tr}(\mathbf{R})^2}{\operatorname{Tr}(\mathbf{R})^2} = \frac{\operatorname{Tr}(\mathbf{R}^2)}{\operatorname{Tr}(\mathbf{R})^2},
	\label{asymt_ch}
\end{equation}
where the rightmost equality comes from the properties of Gaussian vectors \cite[Lemma B.14]{bjornson_massive_2017}.
This result %given using the hypothesis \CN~only
only depends on the covariance matrix $\mathbf{R}$, from which the influences of antenna array topology and propagation conditions are not explicitly identified. Moreover, small-scale and large-scale phenomena are not easily separated either. 
{In this paper, \eqref{non_asymt_ch} is studied using a {physical} channel model that leads to much more interpretable results.}
%This approach has also been previously used recently in several papers like \cite{martinez_massive_2016} where the authors use a measured channel and perform statistics on the received power to show the influence of channel hardening. The same method is used in the papers \cite{ngo_no_2017} and \cite{bjornson_massive_2017} which consider only a $N_t \times 1$ channel model. The authors define channel hardening by the behavior of this metric for asymptotically large antenna arrays (\ref{asymt_ch}).

%\subsection{Statistical model}
\label{model_hypotheses}
{\noindent {\bf Assumptions on the channel model.}} The multipath channel model described in Section
\ref{channel_model} rel\mc{ies} on several parameters governed by some statistical laws. Our aim is to provide an analytical analysis of $CV$ while relying on the
weakest possible set of assumptions on the channel model. Hence, we will consider that: 

%\mc{Let us introduce the main assumptions used on the ray-based} multipath channel model discussed
%in Section \ref{sec_model}. \mc{Such model relies on several parameters that should be governed by
%	some given statistical laws. Our aim is to provide an analytical analysis of $CV$ while relying on the
%	weakest possible set of assumptions on the channel model. Hence, we will} consider that: 
\begin{itemize}
	\item For each ray, gain, DoD and DoA are independent.
	\item $\operatorname{arg}(c_p) \sim \mathcal{U}[0, 2\pi]$ i.i.d.
	\item $\vec{u}_{tx, p}$ and $\vec{u}_{rx, p}$ are i.i.d. with distributions $\mathcal{D}_{tx}$ and $\mathcal{D}_{rx}$.
\end{itemize}
The first hypothesis is widely used and simply say\mc{s} that no formal relation exists between the gain and the DoD/DoA of each ray. %can easily be justified. %They are widely used in physical channel models.% and observed in various channel measurements \cite{saleh_statistical_1987}.
The second one raisonnably indicates that each propagated path experiences independent phase rotation without any predominant angle.
The last one assumes that all the rays come from independent directions, with the same distribution (distributions $\mathcal{D}_{tx}$ at the emitter, $\mathcal{D}_{rx}$ at the receiver).

It has \mc{indeed} been observed \mc{through} several measurement campaigns that rays can be grouped into clusters \cite{saleh_statistical_1987, wu_60-ghz_2017}. Considering the limited angular resolution of {finite-size} antenna arrays, {it is possible to} approximate all rays of the same cluster as a unique {ray} {without harming a lot the channel description accuracy \cite{magoarou_parametric_2017}.} It then makes sense to assume that this last hypothesis is valid for the main DoDs and DoAs of the clusters.
 
%We only consider small scale fading due to multipath. We won't study the slowly varying shadowing, path-loss and absorption phenomenons. As a consequence in the last hypothesis we consider $\Vert \mathbf{c} \Vert$ constant on the time frame under study.

%\subsection{Simulations}

\label{chap_simu}
{\noindent {\bf Simulations.}} A preliminary assessment of the coefficient of variation is computed
\mc{through} Monte-Carlo simulations \mc{of} \eqref{non_asymt_ch} using uniform linear arrays (ULA) with
inter-antenna spacing of $\frac{\lambda}{2}$ at both the transmitter and receiver and \mc{taking} a growing number of antennas. {A total of $P \in \{2,4,5,6\}$ paths were randomly generated with Complex Gaussian gains $c_p\sim \mathcal{CN}(0,1)$, uniform DoDs $\vec{u}_{tx, p} \sim \mathcal{U}_{\mathcal{S}_2}$ and DoAs $\vec{u}_{rx, p} \sim \mathcal{U}_{\mathcal{S}_2}$.}
{%Simulation parameters are given in Table \ref{sim_parameters}. The channel gains $c_p$ are independent complex Gaussian distributed. DoDs and DoAs are uniformly distributed over the unit sphere (Distribution $\mathcal{U}_{\mathcal{S}_2}$). 
}

{
%\begin{table}[!t]
%% increase table row spacing, adjust to taste
%\renewcommand{\arraystretch}{1.3}
%% if using array.sty, it might be a good idea to tweak the value of
%% \extrarowheight as needed to properly center the text within the cells
%\caption{Simulation parameters}
%\label{sim_parameters}
%\centering
%% Some packages, such as MDW tools, offer better commands for making tables
%% than the plain LaTeX2e tabular which is used here.
%\begin{tabular}{|c||c||c||c||c|}
%\hline
%gain distribution & $D_{tx}$ & $D_{rx}$ & P & antenna array \\
%\hline
%$\mathcal{CN}(0, 1)$ & $\mathcal{U}_{\mathcal{S}_2}$ & $\mathcal{U}_{\mathcal{S}_2}$ & 2, 4, 6, 8 &
%ULA, spacing $\lambda/2$ \\
%\hline
%\end{tabular}
%\end{table}
}
\begin{figure}[!t]
	\centering
	\includegraphics[width=3.2in, trim=0px 5px 0px 30px, clip]{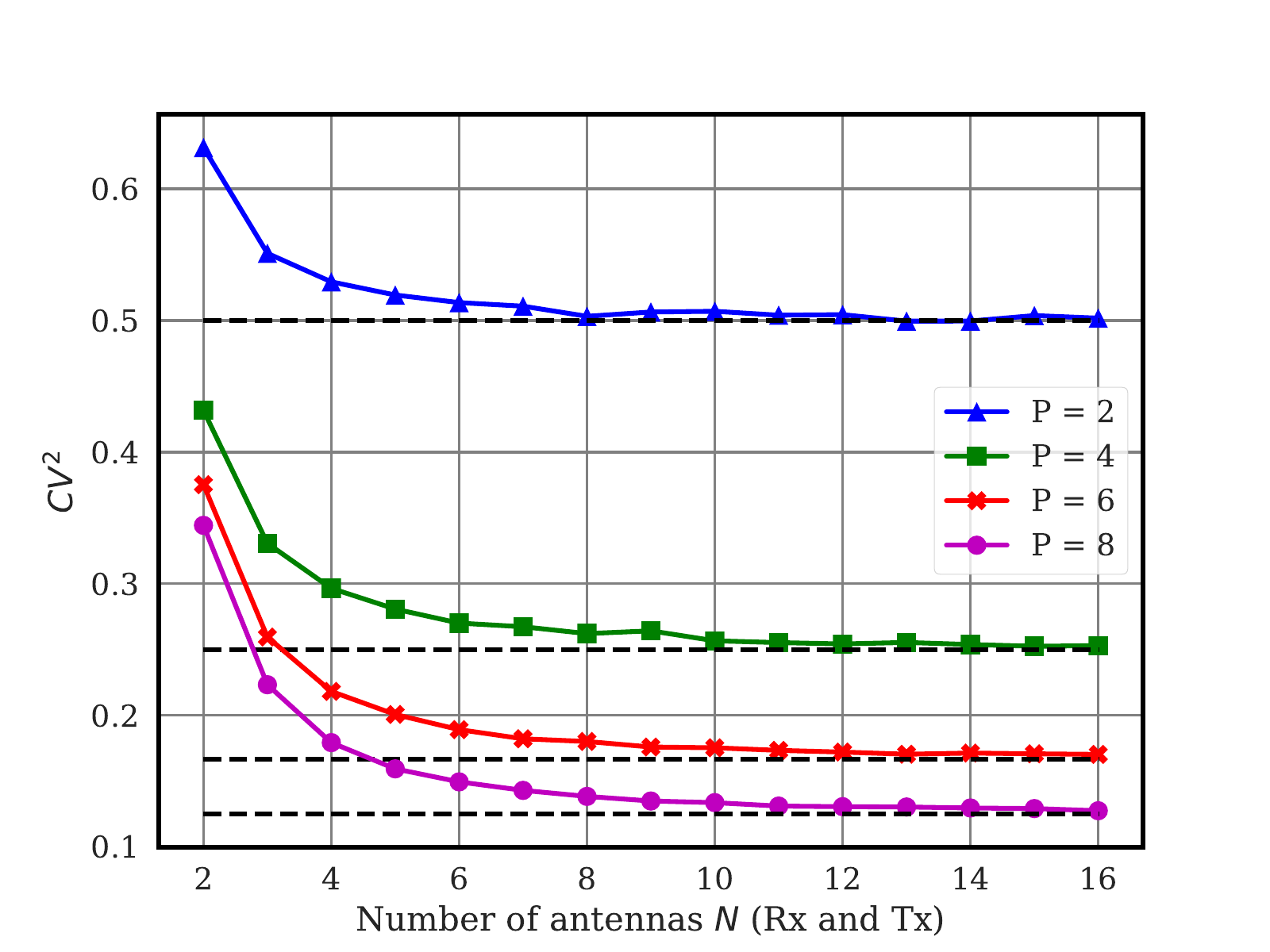}
	\caption{Simulated  $CV^2$ for growing number of rays. Asymptotes are the black dashed lines.}
\label{fig_sim}	
\end{figure}
\mc{Simulation results of $CV$ are reported in} Fig.~\ref{fig_sim} {as a function of
  the number of antennas. \commentaire{, where it is seen}\mc{It is observed} that all curves seem
  to reach an asymptote \mc{around} %defined by
  $1/P$ \mc{for large $N_t$ and $N_r$}. \mc{Hence, the higher the number} of physical paths, the
  harder the channel}.
{%However simulations are not the best tool to show the influence of the different parameters on channel hardening.
} The goal of the next sections is to provide {further} %an easier
interpretation of \mc{such} phenomenon by means of analytical derivations.

\section{Derivation of $CV^2$}
In this section, $CV^2$ is analytically analyzed from \eqref{non_asymt_ch}. 
\commentaire{to provide a better interpretation of channel hardening.}
%\subsection{Expectation}

{\noindent {\bf Expectation of the channel gain.}} {From (\ref{generic_cm}) and (\ref{steering_vect})} the channel gain {$\Vert \mathbf{H} \Vert^2_F = \operatorname{Tr}(\mathbf{H}^H\mathbf{H})$} can be written as
\begin{equation*}
\Vert \mathbf{H} \Vert^2_F = N_t N_r \sum_{p, p'} c_p^* c_{p'}^{\phantom{}}  \gamma_{p, p'},
\label{h_label}
\end{equation*}
where the term $\gamma_{p, p'}$ is given by
\begin{equation*}
\gamma_{p, p'} =  \inner {\mathbf{e_{r}}(\vec{u}_{rx,p})} {\mathbf{e_{r}}(\vec{u}_{rx,p'})}
\inner {\mathbf{e_{t}}(\vec{u}_{tx,p})}{ \mathbf{e_{t}}(\vec{u}_{tx,p'})}^*.
\label{gamma_notation}
\end{equation*}
Using the hypothesis $\operatorname{arg}(c_p) \sim \mathcal{U}[0, 2\pi]$ i.i.d. introduced in the
channel model and $\gamma_{p, p} = 1$, the expectation of the channel gain can further be  expressed as
\begin{equation}
\mathbb{E} \left\{ \Vert \mathbf{H} \Vert_F^2  \right\} 
= N_t N_r \mathbb{E} \left\{ \Vert \mathbf{c} \Vert^2 \right\}.
\label{exp_label_new}
\end{equation}
 \commentaire{In (\ref{exp_label_new})}\mc{Thus the average channel} gain increases
 linearly with $N_r$ and $N_t$\mc{, which is consistent with the expected} beamforming gain $N_t$ and \mc{the
 fact that the received power} linearly depends on $N_r$. %$\Vert \mathbf{c} \Vert^2$ represents the total aggregated power.

\label{coeff_variation}
{\noindent {\bf Coefficient of variation.}} The coefficient of variation $CV$ is derived using the previous hypotheses and  (\ref{exp_label_new}). We introduce\commentaire{ the notations}:
\begin{equation}
\label{notations_geom}
\begin{cases}
\mathcal{E}^2(\mathcal{A}_{tx}, \mathcal{D}_{tx}) = \mathbb{E} \left\{{\vert}\inner {\mathbf{e_{t}}(\vec{u}_{tx,p})}{ \mathbf{e_{t}}^*(\vec{u}_{tx,p'})}{\vert}^2 \right\} \\
\mathcal{E}^2(\mathcal{A}_{rx}, \mathcal{D}_{rx}) = \mathbb{E} \left\{
{\vert}\inner { \mathbf{e_{r}}(\vec{u}_{rx,p})} {\mathbf{e_{r}}(\vec{u}_{rx,p'})} {\vert}^2 \right\}.
\end{cases}
\end{equation}
{These quantities} are the second moments of the inner products of the transmit/receive steering vectors associated \commentaire{with}\mc{to} two distinct rays. They represent the correlation between two rays as observed by the system. They can also be interpreted as the average {inability} of the antenna arrays to discriminate {two} rays given a specific topology and ray distribution. %, the lower the better. 
From such definitions, and based on the {derivations} given in Appendix \ref{appendix}, $CV^2$ {can
  be expressed} as a sum of two terms, 
\begin{equation}
\begin{split}
CV^2 =& \mathcal{E}^2(\mathcal{A}_{tx}, \mathcal{D}_{tx}) \mathcal{E}^2(\mathcal{A}_{rx}, \mathcal{D}_{rx}) \frac{\mathbb{E}\left\{\Vert \mathbf{c}\Vert^4 - \Vert \mathbf{c}\Vert^4_4\right\}}{\mathbb{E}\left\{\Vert \mathbf{c}\Vert^2\right\}^2}\\
+& \frac{\mathbb{V}\text{ar}\left\{\Vert \mathbf{c}\Vert^2\right\}}{\mathbb{E}\left\{\Vert \mathbf{c}\Vert^2\right\}^2}.
\label{CV}
\end{split}
\end{equation}

Note that this result only relies on the assumptions introduced in {section} \ref{channel_model}. The second term can be identified as the contribution of the spatial large-scale phenomena since it simply consists in the coefficient of variation of the previously defined large-scale fading parameter $\Vert \mathbf{c}\Vert^2$ of the channel. \mroy{To allow local channel behavior interpretation, conditioning the statistical model by $\Vert \mathbf{c} \Vert^2$ is required. It results in the cancellation of the large-scale variations contribution of $CV^2$ which reduces to what is called hereafter \it{small-scale fading}.}

%In constrast, the first term can be defined as the spatial small-scale contribution. In the following, a comprehensive analysis of these two terms {with respect to} the main parameters of the ray-based channel model {is carried out}.

\section{Interpretations}

\commentaire{This section highlights the influence of small-scale and large-scale fading on $CV^2$.} 
%\subsection{Large-Scale Fading}
\subsection{Large-scale fading}
\label{large_scale}
%{\noindent {\bf Large-scale fading.}} 
The contribution of large-scale fading in $CV^2$ is basically the coefficient of variation of the
total aggregated power $\Vert \mathbf{c} \Vert^2$ of the rays. \commentaire{This expression matches the second term of the expression (\ref{CV}).}
To better emphasize \mc{its} behavior, {let us} consider a simple example with
independent $\vert c_p \vert^2$ {of mean $\mu$ and variance $ \sigma^2 $. The resulting large scale
  fading term is then
	\begin{equation*}
	\frac{\mathbb{V}\text{ar}\left\{\Vert \mathbf{c}\Vert^2\right\}}{\mathbb{E}\left\{\Vert \mathbf{c}\Vert^2\right\}^2} = \frac{1}{P}\left(
	\frac{\sigma}{\mu}\right)^2.
	\label{toy_example2}
	\end{equation*}}
%\vspace{0.2cm}
%
%\begin{minipage}{0.15\textwidth}
%	\begin{equation*}
%	\begin{split}
%	\mathbb{E}\left\{\vert c_p \vert^2\right\} &= \mu\\ 
%	\operatorname{\mathbb{V}\text{ar}}\left\{\vert c_p \vert^2\right\} &= \sigma^2
%	\end{split}
%	\label{toy_example}
%	\end{equation*}
%\end{minipage}
%\hspace{0.1cm}
%\begin{minipage}[l]{0.1\textwidth}
%	\raggedright
%	resulting in large-scale:
%\end{minipage}
%\hspace{0.1cm}
%\begin{minipage}[c]{0.16\textwidth}
%	\begin{equation*}
%	\frac{\mathbb{V}\text{ar}\left\{\Vert \mathbf{c}\Vert^2\right\}}{\mathbb{E}\left\{\Vert \mathbf{c}\Vert^2\right\}^2} = \frac{1}{P}\left(
%	\frac{\sigma}{\mu}\right)^2.
%	\label{toy_example2}
%	\end{equation*}
%\end{minipage}
%\vspace{0.2cm}

%The resulting large-scale term is given by
It clearly appears that more rays {lead to} reduced large-scale variations. This {stems} %could be expected 
from the fact that any shadowing phenomenon is well averaged over $P$ independent rays, hence becoming almost deterministic in rich scattering environments. %In a general case, it can be well understood that the more rays, the fewer large-scale fading. 
This result explains the floor levels obtained for various $P$ in our previous simulations in Section \ref{chap_simu} and is consistent with the literature on correlated Rayleigh fading channels where high rank correlation matrices provide a stronger channel hardening effect than {low rank} ones \cite{bjornson_massive_2017}. 

%\subsection{Small-Scale Fading}
%{\noindent {\bf Small-scale fading.}}
\subsection{Small-scale fading}
The coefficient of variation \commentaire{computed in Section \ref{coeff_variation} assuming fixed
  large-scale gain $\Vert \mathbf{c} \Vert^2$}\mroy{particularized with the statistical conditional
  model} \mc{can easily be} proven to be:
\begin{equation}
\begin{split}
CV^2_{\Vert \mathbf{c}\Vert^2 } &= \mathcal{E}^2(\mathcal{A}_{tx}, \mathcal{D}_{tx}) \mathcal{E}^2(\mathcal{A}_{rx}, \mathcal{D}_{rx}) \alpha^2(\mathbf{c})\\
\text{where }\alpha^2(\mathbf{c}) &= 1 - \frac{\mathbb{E}_{\mathbf{c}\vert \Vert \mathbf{c}\Vert^2}\left\{\Vert \mathbf{c}\Vert^4_4  \right\}}{\Vert \mathbf{c}\Vert^4}.
\label{CV_2}
\end{split}
\end{equation}
The small-scale fading contribution to $CV^2$ thus consists of a product of the {quantities defined in (\ref{notations_geom}) that depend only on the antenna array topologies ($\mathcal{A}_{tx}$/$\mathcal{A}_{rx}$) and ray distributions ($\mathcal{D}_{tx}$/$\mathcal{D}_{rx}$) multiplied by a propagation conditions factor $\alpha^2(\mathbf{c})$ that depends only on the statistics of the ray powers $\mathbf{c}$.} %$\mathcal{E}$ is the same at transmitter and receiver and only depends on the corresponding antenna array topology ($\mathcal{A}_{tx}$/$\mathcal{A}_{rx}$) and ray distribution ($\mathcal{D}_{tx}$/$\mathcal{D}_{rx}$). The propagation factor $\alpha^2(\mathbf{c})$ is only driven by the statistics of the ray powers $\mathbf{c}$.
%In the particular case of the simulations performed in \ref{chap_simu}, the propagation term is equal to $1-1/P$ \stpa{(upper-bound)}. 

%\subsection{Numerical evaluation of $\mathcal{E}$}
\noindent {\bf {Ray correlations.}} 
This {paragraph focuses} on {the quantity $\mathcal{E}^2(\mathcal{A}_{tx},~\mathcal{D}_{tx})$} {(the study is done only at the emitter, the obtained results {being equally valid} at the receiver). %Function $\mathcal{E}^2$ is the second moment of the sum of non-independent phases
Eq. \eqref{notations_geom} yields}
\begin{equation*}
\mathcal{E}^2(\mathcal{A}_{tx},~\mathcal{D}_{tx}) = \frac{1}{{N_t^2}}\mathbb{E} \left\{\left| \sum_{i = 1}^{N_t} \operatorname{e}^{2 j \pi \frac{\vec{a}_{tx,i} \cdotp (\vec{u}_{tx, p} - \vec{u}_{tx, p'}) }{\lambda}} \right|^2 \right\}.
\label{geom_detail}
\end{equation*}
\mc{A well-known situation is when the inner sum involves exponentials of independent uniformly
  distributed phases and hence corresponds to a random walk with $N_t$ steps of unit length. The
  above expectation then consists in the second moment of a Rayleigh distribution and
  $\mathcal{E}^2(\mathcal{A}_{tx},~\mathcal{D}_{tx}) = \frac{1}{{N_t}}$. A necessary condition to
  such a case is to have (at least) a half wavelength antenna spacing $\Delta d$ to ensure that
  phases are spread over $[0,\, 2\pi]$. On the other hand, phase independences are expected to occur
  for asymptotically large $\Delta d$. It is however shown hereafter that such assumption turns
  out to be valid for much more raisonnable value of $\Delta d$.}

\commentaire{For asymptotically large antenna spacing $\Delta d$, the phases of each term of the inner sum are independent and identically distributed. In that case, the {inner sum} is a random walk with $N_t$ steps of unit length, whose modulus follows a Rayleigh distribution of parameter $\tfrac{N_t}{2}$. The expectation is then the corresponding second moment, which is equal to $N_t$. This yields $\mathcal{E}^2(\mathcal{A}_{tx}, \mathcal{D}_{tx})\xrightarrow{\Delta d\rightarrow \infty} 1/N_t$. }

Numerical evaluations of $\mathcal{E}^2$ are performed versus $\Delta d$ (Fig. \ref{fig_sim_D}), and
versus $N_t$ (Fig. \ref{fig_sim_N}). Uniformly distributed rays over the 3D unit sphere
($\mathcal{D}_{tx} = \mathcal{D}_{rx} = \mathcal{U}_{\mathcal{S}_2}$) and Uniform Linear, Circular
and Planar Arrays (ULA, UCA and UPA) are considered. As a reminder, the {smaller}
$\mathcal{E}(\mathcal{A}_{tx}, \mathcal{D}_{tx})$ the better {the channel hardening}.  {In
  Fig. \ref{fig_sim_D}}, $\mathcal{E}^2$ reaches the asymptote $1/N_t$ for all array {types} with
$\Delta d = \tfrac{\lambda}{2}$ and remains almost constant for larger $\Delta
d$.
Fig. \ref{fig_sim_N} shows that $\mathcal{E}^2$ merely follows the $1/N_t$ law whatever the array
type. \mroy{We thus conclude that \mc{the independent uniform phases situation} discussed above is a
  sufficient model for any array topology given that $\Delta d \ge \tfrac{\lambda}{2}$.} It is
therefore assumed in the sequel that,

\commentaire{Smaller spacings imply a more
  unstable channel gain whereas greater ones lead
  to} %after the half-wavelength channel hardening is always close to optimal for the given configuration. Considering at least half-wavelength antenna spacing in this paper, we assume in the following
\begin{equation*}
	\mathcal{E}^2(\mathcal{A}_{tx}, \mathcal{U}_{\mathcal{S}_2}) {\approx} 1/N_t, \quad \mathcal{E}^2(\mathcal{A}_{rx}, \mathcal{U}_{\mathcal{S}_2}) {\approx} 1/N_r.
\end{equation*}

\commentaire{Considering at least half-wavelength antenna spacing in this paper, we assume in the following $\mathcal{E}^2(\mathcal{A}_{tx}, \mathcal{U}_{\mathcal{S}_2})= 1/N_t $  at the emitter ($1/N_r$ at the receiver, resp.) for any \commentaire{antenna} array topology.}

\begin{figure}[!t]
	\centering
	\includegraphics[width=3.2in, trim=0px 5px 0px 30px, clip]{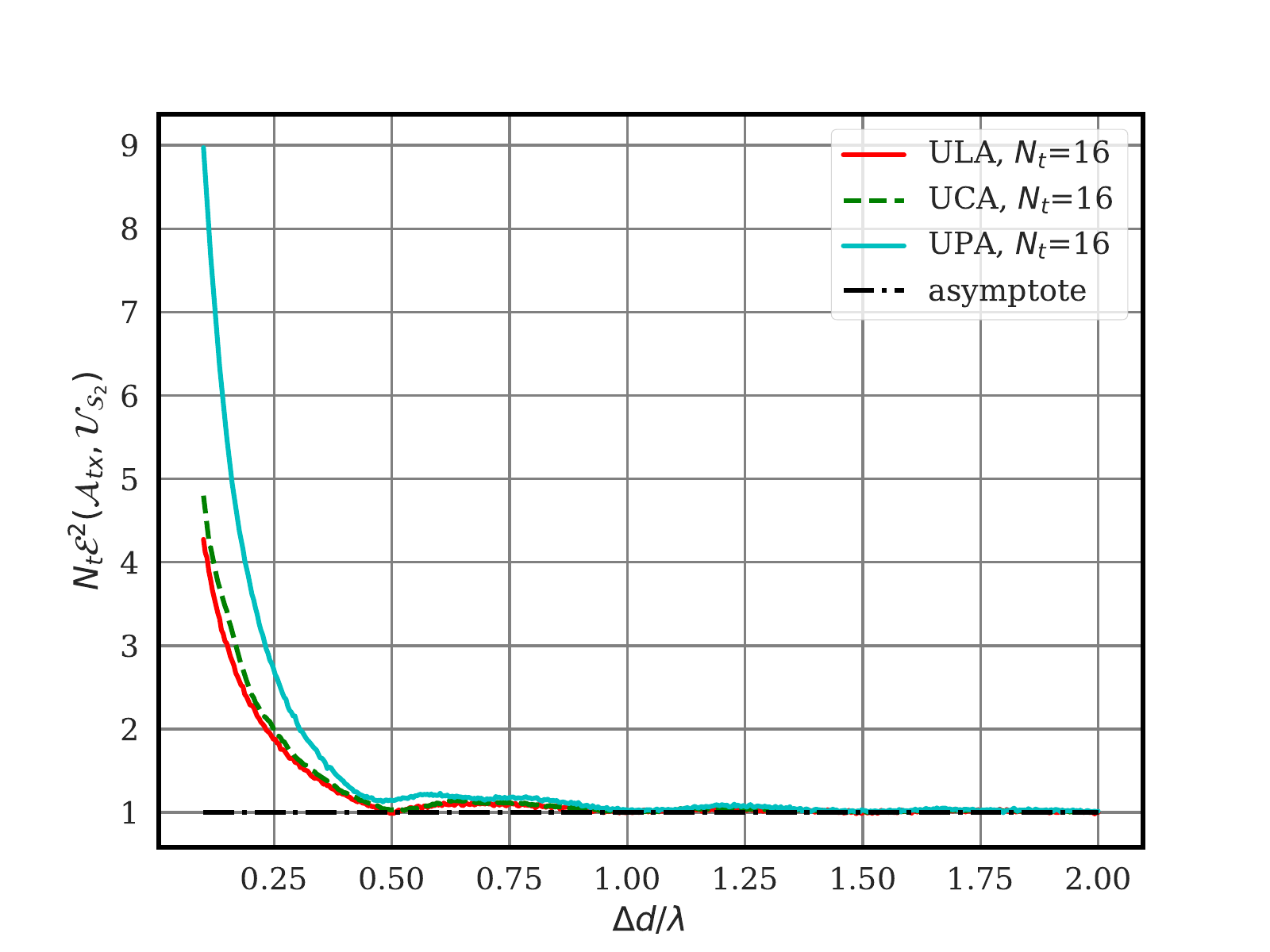}
	\caption{Numerical evaluation of $\mathcal{E}(\mathcal{A}_{tx},
          \mathcal{U}_{\mathcal{S}_2})$ for various array \mc{types} and increasing antenna spacing
          $\Delta d$.\commentaire{The lower the better.} The values are normalized so the asymptote is 1. \commentaire{We can spot that for the ULA and the UPA, the first minimum is at {half the} wavelength.}}
	\label{fig_sim_D}
\end{figure}
\begin{figure}[!t]
	\centering
	\includegraphics[width=3.2in, trim=0px 5px 0px 30px, clip]{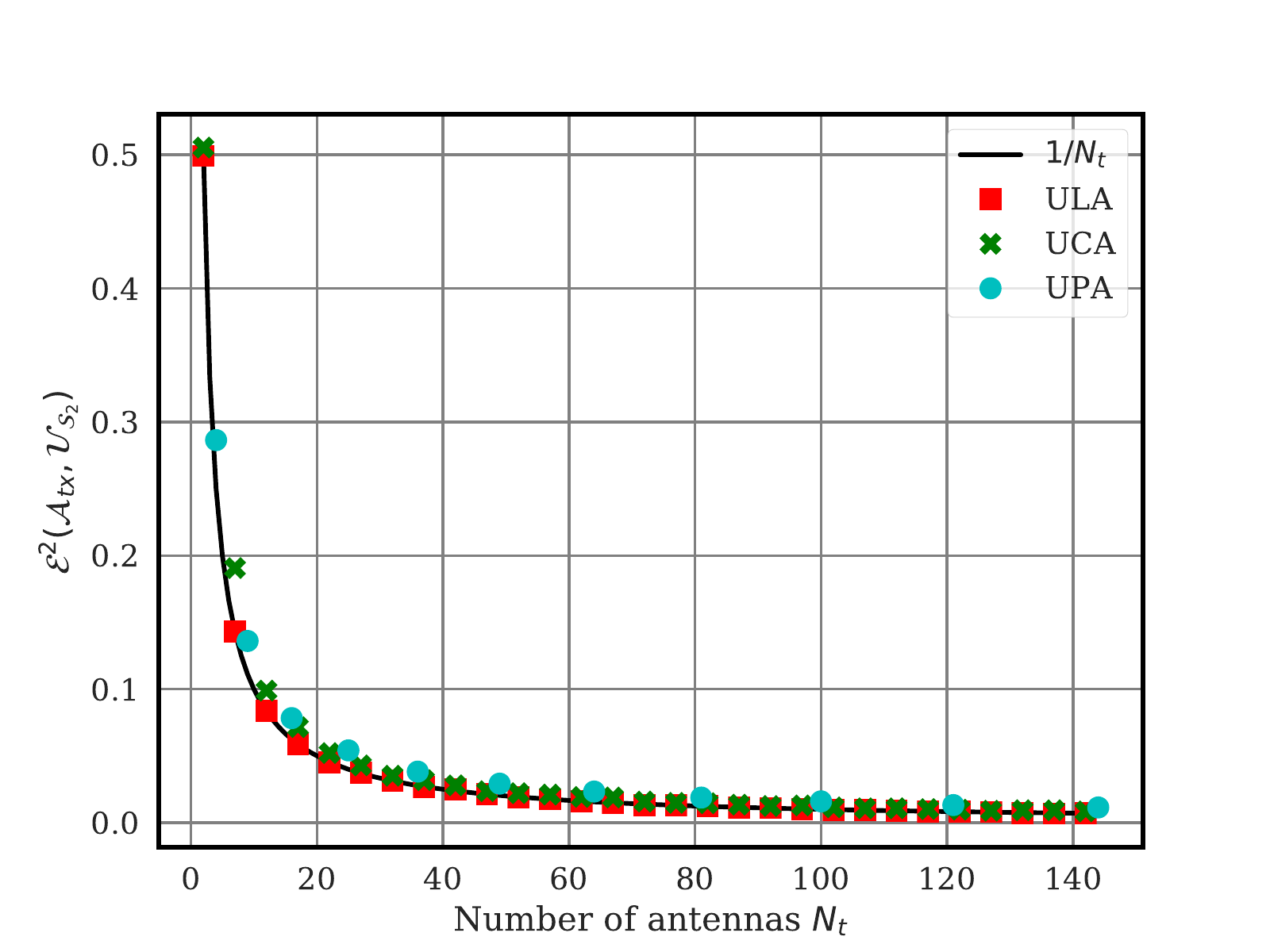}
	\caption{Numerical evaluation of $\mathcal{E}(\mathcal{A}_{tx}, \mathcal{U}_{\mathcal{S}_2})$ for various antenna arrays at the half wavelength. The lower, the better. }
	\label{fig_sim_N}
\end{figure}
%\subsection{Comparison with the simulations}

{\noindent {\bf {Propagation conditions.}}} 
\mc{It is now interesting to point out that} the propagation factor $\alpha(\mathbf{c})$ introduced in  (\ref{CV_2}) \commentaire{only depends on the ray powers distribution.}
\mroy{is bounded by}
\begin{equation}
0\leq \alpha^2(\mathbf{c})\leq 1 - 1/P.
\label{prop_bounds}
\end{equation}
Those bounds are deduced from the following inequality:
\begin{equation}
\Vert \mathbf{c} \Vert_2^4/P \leq \Vert \mathbf{c} \Vert_4^4 \leq \Vert \mathbf{c} \Vert^4_2 .
\label{convex_ineq}
\end{equation}
The right inequality comes from the convexity of the square function. Equality is achieved when there is only one contributing ray, \mc{i.e. no multipath occurs}. In that case $CV^2_{\Vert \mathbf{c}\Vert^2 } = 0$ and the MIMO channel power is deterministic. 
The left part in (\ref{convex_ineq}) is given by {H\"older's} inequality. {Equality} is achieved when there are $P$ rays {of equal power}. 
Then, {taking the expectation on each member in (\ref{convex_ineq}) yields} (\ref{prop_bounds}).% that $\alpha(\mathbf{c})$ is bounded.

{In contrast to the} large-scale fading, more rays {lead to} more small-scale fluctuations. It is indeed well known that a {richer} scattering environment {increases} small-scale fading.

{\noindent {\bf Comparison with the simulations.}} Based on the general formula given in (\ref{CV}), on the interpretations and evaluations of its terms, we can derive the expression of channel hardening for the illustrating simulations of Section \ref{chap_simu}:
\begin{equation*}
	CV^2_{\text{illustration}} = \frac{1}{N_tN_r}(1-1/P) + 1/P.
\end{equation*}
Simulation and approximated formula are compared in Fig. \ref{fig_sim_verif} \mc{in which} small-scale
and large-scale contributions are easily evidenced, as intuitively expected from simulations of Fig. \ref{fig_sim}.

\begin{figure}[!t]
	\centering
	\includegraphics[width=3.2in, trim=0px 5px 0px 30px, clip]{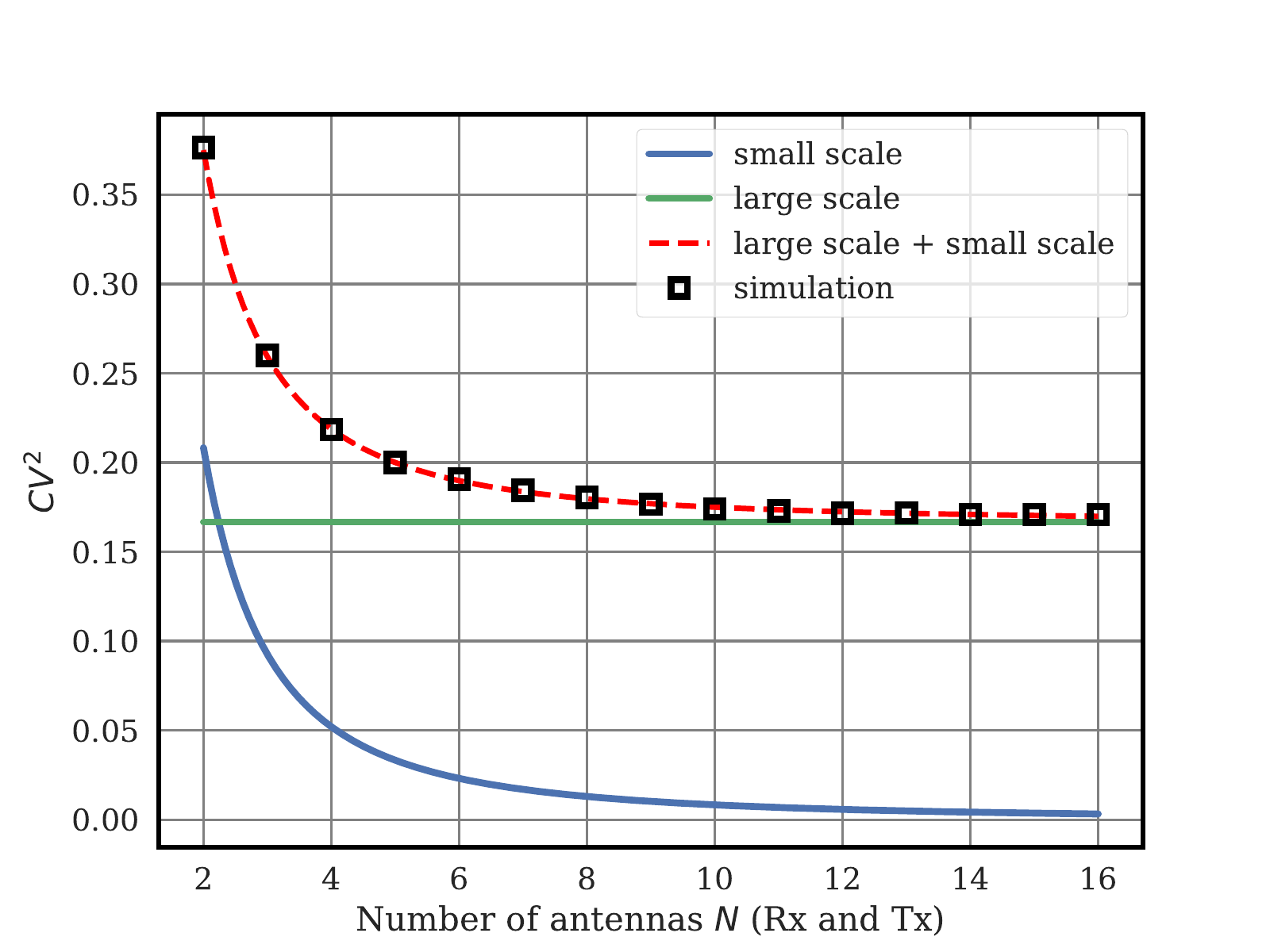}
	\caption{Comparison between  (\ref{CV}) and simulated  $CV^2$. Uniform distribution of DoDs and DoAs over the unit sphere and complex Gaussian gains. \commentaire{Large-scale and small-scale influences are evidenced.}}
	\label{fig_sim_verif}
\end{figure}

{\noindent {\bf Comparison with the Gaussian i.i.d. model.}} This model assumes a rich scattering environment.  Using (\ref{asymt_ch}) with $\mathbf{R} = \mathbf{I}$:
\begin{equation*}
CV^2_{iid} = \frac{1}{N_tN_r}.
\label{cs_iid}
\end{equation*}
Using the realistic model in a rich scattering environment, the large-scale part of (\ref{CV}) vanishes leading to a deterministic $\Vert \mathbf{c} \Vert^2$ and small-scale variations reach the upper bound of (\ref{prop_bounds}). This yields the limit   
\begin{equation}
CV^2 \xrightarrow{P\rightarrow \infty} CV^2_{iid}
\label{cs_convergence}
\end{equation}
which is coherent with the interpretation of the \commentaire{Gaussian i.i.d. }model.

\section{Conclusion}
%MIMO channel advantages are twofold. Spatial multiplexing techniques have been widely studied over the previous years but the "channel hardening" effect wasn't clearly defined until recently \cite{ngo_no_2017, bjornson_massive_2017}. However the results are difficult to interpret physically because purely stochastic models were used. 

In this paper, previous studies on channel hardening \mroy{have been} extended using a physics-based model. We \mroy{have separated} influences of antenna array topologies and propagation characteristics on the channel hardening phenomenon. Large-scale and small-scale contributions to channel variations have been evidenced. Essentially, this paper provides a general framework to study channel hardening using accurate propagation models.

To illustrate the overall behavior of channel hardening, this framework \mroy{have been} used with generic model parameters and hypotheses. The scaling laws evidenced for simpler channel models are conserved provided the antennas are spaced by at least half a wavelength. The results are consistent with state of the art and provide further insights on the influence of array topology and propagation on channel hardening. The proposed expression  can easily be exploited with various propagation environments and array topologies to provide a more precise understanding of the phenomenon compared to classical channel descriptions {based on \mroy{Rayleigh fading models}.}


\section*{Acknowledgment}

{
\footnotesize
This work has been performed in the framework of the Horizon 2020 project ONE5G (ICT-760809) receiving funds from the European Union. The authors would like to acknowledge the contributions of their colleagues in the project, although the views expressed in this contribution are those of the authors and do not necessarily represent the project.
}

\appendices
\section{Coefficient of variation { (\ref{CV})}}
\label{appendix}
For the sake of simplicity, an intermediary matrix $\mathbf{A}$ is introduced. It is defined by 
\begin{equation*}[\mathbf{A}]_{p, p'} =\left\{\begin{array}{ll}
2\vert \gamma_{p, p'} \vert \operatorname{cos}(\phi_{p, p'}) & \text{if } p \neq p' \\
1 & \text{if } p = p' \\
\end{array}\right.
\end{equation*}
with $\phi_{p, p'}=\text{arg}(c_p^* c_{p'}^{\phantom{}} \gamma_{p, p'})$ the whole channel phase dependence.
$\Vert \mathbf{H} \Vert^2_F$ can be written using a quadratic form with vector $\mathbf{c}$ and matrix $\mathbf{A}$, which can be decomposed into two terms $\mathbf{I}$ (identity) and $\mathbf{J}$

\begin{equation*}
\frac{\Vert \mathbf{H} \Vert^2_F}{N_t N_r} =  \mathbf{c}^T  \mathbf{Ac} = \mathbf{c}^T  \mathbf{c} + \mathbf{c}^T  \mathbf{Jc}
\label{quadratic_form}
\end{equation*}
where $\mathbf{J} = \mathbf{A}-\mathbf{I}$. $\mathbb{E} \left\{ \mathbf{J} \right\} = \mathbf{0}$ so:
\begin{equation*}
\frac{\mathbb{E} \left\{ \Vert \mathbf{H} \Vert_F^4 \right\}}{(N_tN_r)^2} = \mathbb{E} \left\{ \Vert \mathbf{c} \Vert^4 \right\} + \mathbb{E} \left\{ (\mathbf{c}^T \mathbf{J} \mathbf{c})^2 \right\}.
\label{sec_order_new}
\end{equation*}
The ray independence properties yields the following weighted sum of coupled ray powers
\begin{equation*}
\mathbb{E} \left\{ (\mathbf{c}^T \mathbf{J} \mathbf{c})^2 \right\} =  \sum_{p\neq p'}
\mathbb{E} \left\{ \vert c_{p^{\phantom{}}} \vert^2 \vert c_{p'} \vert^2 \right\} \mathbb{E} \left\{ [\mathbf{J}]_{p, p'}^2
\right\}.
\label{sec_order_mp2}
\end{equation*}

Considering i.i.d. rays, all the weights $\mathbb{E} \left\{ [\mathbf{J}]_{p, p'}^2\right\}$ are identical. Using the weights notations introduced in (\ref{notations_geom}) and the definition of the 4-norm yields the second order moment $\mathbb{E} \left\{ \Vert \mathbf{H} \Vert_F^4 \right\}$. With the expectation (\ref{exp_label_new}) we derive the result (\ref{CV}).

\bibliographystyle{IEEEtran}
\bibliography{biblio_zotero_reduced}
%\begin{thebibliography}{1}
%\end{thebibliography}

\end{document}